\documentclass[aps,prc,dlspace,twocolumn,a4paper]{revtex4}
\usepackage{amsfonts}
\usepackage{epsfig}
\usepackage{graphics}
\usepackage{color}
\newcommand{\beq}{\begin{equation}}
\newcommand{\eeq}{\end{equation}}
\newcommand{\bqa}{\begin{eqnarray}}
\newcommand{\eqa}{\end{eqnarray}}
\newcommand{\jsi}{J / \psi}
\newcommand{\chic}{\chi_{c}}

\newcommand{\eps}{\epsilon_s}

\newcommand{\jpsi}{J / \psi}

\newcommand{\pt}{p_{{}_T}}
\newcommand{\ptmin}{p_{{}_T}^{\mbox{min}}}
\newcommand{\ptmax}{p_{{}_T}^{\mbox{max}}}
\newcommand{\spt}{S(p_{{}_T})}
\newcommand{\sinpt}{ \langle S(p_{{}_T})\rangle}

\newcommand{\sdir}{\langle S^{{}^{\mbox{dir}}} \rangle}

\newcommand{\msbar}{\overline{\mbox{\rm MS}}}

\newcommand{\roots}{\sqrt{s}}
\newcommand{\fm}{{\rm fm}}
\newcommand{\gev}{{\rm GeV}}

\newcommand{\lambdamsbar}{\Lambda_{\overline{\rm MS}}}
\begin{document}
\title{Charmonium suppression at RHIC: Signature of a strongly-interacting QGP, not a weakly interacting}
\author{Binoy Krishna Patra$^1$}
\email{binoyfph@iitr.ernet.in}
\author{Vineet Agotiya$^1$}
\affiliation{$^1$ {\small Department of Physics, Indian Institute of 
Technology Roorkee, Roorkee-247 667, India}}

\begin{abstract}
Following a recent work on equation of state for strongly interacting 
quark-gluon plasma~\cite{banscqgp}, we revisited the equation of state 
by incorporating the nonperturbative effects in the deconfined 
plasma phase. Our results on thermodynamic observables
{\em viz.} pressure, energy density, speed of sound etc. nicely fit with
the lattice equation of state for gluon, massless and as
well {\em massive} flavored plasma. Motivated by this agreement with
lattice results, we have employed our equation of state to estimate the
quarkonium suppression in an expanding, dissipative strongly interacting
QGP produced in relativistic heavy-ion collisions and our prediction
matches exactly with the recent PHENIX data on the centrality
dependence of $J/\psi$ suppression in Au+Au collisions at BNL RHIC.
We have also predicted for the $\Upsilon$
suppression in Pb+Pb collisions at LHC energy
which could be tested in the ALICE experiments at CERN LHC.

\end{abstract}
\maketitle


\section{Introduction}
Quantum chromodynamics (QCD) at high temperature is believed to be in 
quark gluon plasma (QGP) phase, whereby color charges are screened rather 
than confined. Asymptotic freedom of non-abelian gauge theories
insures that for high enough temperature $T\gg {\Lambda_{QCD}}$, QGP 
is weakly coupled with dressed quarks and gluons behaving as quasi-particles
near the ideal gas limit.

In relativistic heavy ion collisions at RHIC, two novel phenomena - cone
and ridge, not present in {\it pp} or d+Au collisions, were 
observed~\cite{cone-rdige}. Quark-gluon plasma which is produced 
in heavy ion collisions is not an ideal gas of quarks and gluons, it is 
rather a liquid having very low shear viscosity to entropy density ($\eta/s$)
ratio~\cite{star,vis1,shur,son}.
This strongly suggest that QGP may lie in the non-perturbative domain of 
QCD which is very hard to address both analytically and computationally.
Similar conclusion about QGP have been reached from recent lattice studies
which predict that the equation of state (EoS) is interacting even at 
$T\sim 4 T_c$~\cite{leos,cheng,karsch,gavai}.
Why QGP is strongly interacting in this temperature range and what is 
its meaning are not very well understood till the date. Since then, attempts 
have been made to understand strongly interacting nature of QGP and its small 
$\eta/s$ ratio, argued from large elliptic flow observed
in RHIC data. Similar conclusions  about the 
near perfect fluidity of QGP have been reached from the AdS/CFT studies
~\cite{son}, spectral functions and transport coefficients in lattice 
QCD~\cite{satz} and studies based on classical strongly coupled plasmas
~\cite{shur1,shur2}.

There are several attempts to model the EoS of such strongly-interacting 
matter {\em viz.} bag model, confinement models, quasi-particle models, 
strongly interacting quark gluon plasma\cite{sqgp,banscqgp} etc.  
In the bag model \cite{rh.1}, QGP 
is treated as a big hadron with large number of
partons interacting weakly but confined by the bag wall.
Further inclusions of glue balls or hadrons improve
the predictions. The confinement models are the extension of bag model with
smooth potential like Cornell potential \cite{qgp_cornell} etc. with a better
predictions. There is an interesting attempt \cite{arnold,zhai} to
determine the EoS of interacting quarks
and gluons upto $O$($g^5$) which 
has been further improved upon \cite{klaine,klaine1} by
incorporating the contributions from the nonperturbative
scales viz. $gT$ and $g^2 T$ up to O($g^6\ln(1/g)$)\cite{krebh}. 
On the other hand, a semiclassical approach aimed to study the 
bulk properties of QGP automatically incorporate hard thermal loop (HTL) 
effects\cite{pfk,pfk1} where the nonperturbative features
manifested as effective mean color fields~\cite{jpb4}.

There are different versions of quasi-particle models~\cite{pe.1} where
equation of state was derived with temperature dependent parton masses 
and bag constant\cite{lh.1,pe.2}, with effective degrees 
of freedom~\cite{s.1}, etc.   
All of them claim to explain lattice results, either by adjusting free 
parameters in the model or by taking lattice data on one of the 
thermodynamic quantity as an input and 
predicting other quantities. However, physical 
picture of quasi-particle model and the origin of various temperature 
dependent quantities are not clear yet~\cite{rh.2}.   
In strongly interacting QGP~\cite{sqgp1,sqgp2,sqgp3}, one considers 
all possible hadrons even 
at $T\,>\,T_c$ and try to explain non-ideal behavior of QGP near $T_c$. 
Recently, an equation of state for strongly-coupled plasma has been 
inferred by 
utilizing the understanding from strongly coupled QED plasma~\cite{ba_cor.1} which 
fits lattice data well. To be honest, deep and 
comprehensive understanding is still missing and future investigations 
(both theoretical and forthcoming experiments at LHC) may throw more light 
on this very exciting and complex issue.

A suppressed yield of quarkonium in the dilepton spectrum,
measured in experiments~\cite{jpsi_sps,exp} was proposed as a signature of QGP
formation. Understanding the experimental data, however, turned out to 
be more complicated because the suppression pattern seen is not only 
due to the hot medium effects of screening~\cite{matsui,satz}, but more 
likely due to interplay of different interactions {\em viz.} cold nuclear 
matter~\cite{cold} effect, gluonic dissociation~\cite{gdisso} as well 
as those of recombination~\cite{reco} of $ c \bar c $ pairs etc. 
In order to disentangle these different effects we must know the properties of 
quarkonium in medium and their dissociation. 
There are two main lines of theoretical studies to determine quarkonium 
spectral functions at finite temperature: potential models~\cite{pnrqcd}-\cite{wong}
which have been widely used to study quarkonium, but 
their applicability at finite temperature is still under scrutiny 
and lattice QCD~\cite{satz,lattice} which provides the most 
straightforward way to determine 
spectral functions, but the results suffer from discretization effects and 
statistical errors, and thus are still inconclusive. 

The short and intermediate distance properties of the heavy quark 
interaction are important for the understanding of in-medium 
modifications of the heavy quark bound states
whereas the large distance behavior plays 
a crucial role in understanding the bulk properties of the QCD plasma 
phase~\cite{kaczmarek}. In the study of bulk thermodynamic quantities 
{\em e.g.} pressure, energy density etc., deviations 
from perturbative calculations and ideal gas behavior 
are found at temperatures much larger 
than the deconfinement temperature which is further supported by
robust collective flows, strong jet and charm quenching,
and charm flow, in RHIC data.
This calls for a quantitative
nonperturbative calculations. Since the phase transition in full QCD
appears as a crossover rather than a true phase transition~\cite{phaseT}, 
it is reasonable to assume that the string tension does not vanish
abruptly above $T_c$. So one should study its effects on the behavior 
of quarkonia in a hot QCD medium. Recently we~\cite{prc-vineet,chic-vineet} 
had considered this potentially interesting
issue by correcting the full Cornell potential with a dielectric
function embodying the effects of the deconfined medium and
not only its Coulomb part as usually done in the literature. This
led to a long-range Coulomb potential
in addition to the usual Debye-screened form. With 
such an effective potential, we had investigated the
effects of perturbative as well as nonperturbative effects 
in QGP on the dissociation of different quarkonium 
states~\cite{prc-vineet,chic-vineet}.

Now let us consider a central collision in a nucleus-nucleus collision,
which results in formation of QGP at initial time $\tau_i$. We assume the
plasma to cool, according to Bjorken's boost invariant longitudinal 
expansion. There are many attempts~\cite{chu,dpal,compare} to incorporate
the plasma expansion dynamics to study charmonium suppression in 
relativistic nuclear collisions. However, some important
points were not addressed in their works~\cite{chu,dpal,compare}
to quantify the suppression in an expanding system: the viscous forces 
in the energy-momentum tensor and the proper EoS. The effects of 
dissipative forces were not
included in hydrodynamic expansion which cause the plasma to 
expand slowly and results in more suppression. The equation of state
employed in their works was either ideal or bag model EoS 
used to calculate two vital factors - screening energy 
density, $\epsilon_s$ of the system (corresponding to dissociation 
temperature) and the time, $\tau_s$ elapsed by the system to reach
at $\epsilon_s$, through expansion. Since the matter formed at RHIC 
is far from its ideal limit even at $T \geq T_c$ so the ideal or bag model
equation of state is not
reliable to study the suppression of charmonium yields in an expanding
strongly interacting system.

Recently we~\cite{eurjc-vineet} had studied the survival of charmonium 
states in a dissipative strongly interacting QGP 
where the suppression of $J/\psi$'s due to potential screening in the 
deconfined medium at relativistic nuclear collisions is two step 
learning. First one is the understanding of dissociation in
a static thermal medium for which knowledge of medium dependence of heavy quark potential is very much needed. Second one is the generalization 
of dissociation in an expanding medium where a equation of state 
plays the major role as an
input. In \cite{eurjc-vineet}, we used the equation of state 
for QGP, in analogy with strongly-coupled plasma
where hadrons exist for $T<T_c$ and go to 
plasma of quarks and gluons (QGP) for $T>T_c$ and there is no hadrons or glue balls
because confinement interactions due to QCD vacuum was assumed to be
melted~\cite{banscqgp}, although Zahed and Shuryak~\cite{sqgp2} suggested
that QGP at temperatures up to few $T_c$ supports weakly bound mesonic 
states. So the only interaction present in the deconfined plasma phase is 
Coulomb interaction and hence plasma 
parameter ($\Gamma$) becomes the ratio of average Coulomb potential
energy (=$\frac{4}{3}~\frac{\alpha_s}{r_{av}}$) to average kintic 
energy ($\sim T$). Finally, an expression for equation of state was 
obtained as a function of plasma parameter, $\Gamma$~\cite{banscqgp}.

Using the abovementioned equation of state, we~\cite{eurjc-vineet} got 
a better agreement with the PHENIX experimental results~\cite{expt}
compared to earlier works\cite{dpal} but still it lacks
complete agreement. The reason of disagreement may
be two fold. This could be either due to the arbitrariness in the definition
of dissociation criteria or due to the equation of state 
which is not fully compatible with the nonperturbative nature of QGP. 
In the present work, we revisited the abovementioned 
equation of state~\cite{banscqgp} by incorporating nonperturbative
effects in the plasma parameter around which the equation of state is expanded.

As discussed above, the existence of nonperturbative interactions even 
at $T\ge T_c$ indicates that the string tension may not vanish abruptly 
at $T_c$, so potential in the deconfined phase could have
a nonvanishing confining (string) term, in addition to the Coulomb 
term~\cite{prc-vineet} unlike Coulomb interaction alone in the
aforesaid model~\cite{banscqgp}. This is the central theme of our work
where EoS has been obtained by retaining both terms in 
the potential and then calculate the thermodynamic 
variables {\em viz} pressure, energy density,
speed of sound etc. Our results match nicely with the lattice results 
of gluon \cite{leos}, 2-flavor (massless) as well as 3-flavor (massless) 
QGP \cite{ka.2}. There is also an agreement with (2+1) (two massless and 
one is massive) and 4 flavoured lattice results too.  
Motivated by the agreement with lattice results, we employ our equation 
of state to study the $J/\psi$ suppression
in an expanding plasma in the presence of viscous forces with the universal ratio $\eta/s=1/4\pi$~\cite{son}. We 
have found complete agreement with the PHENIX results on the centrality
dependence of $J/\psi$ suppression
at RHIC energy. Further we have predicted the $\Upsilon$ suppression which could be the 
potential candidate for the LHC experiment where $J/\psi$ suppression
could be marred by enhancement due to recombination of abundant $c \bar c$ pairs.

The paper is organized as follows. In Sec.II.A, we briefly discuss
our recent work on medium modified potential. Using this effective 
potential, we have then developed the equation of state for 
strongly interacting matter and have shown our results on pressure,
energy density and speed of sound etc. along with the lattice data in 
Sec.II.B. In Sec.III.A, we have employed the aforesaid equation of state
to study boost-invariant (1+1) dimensional longitudinal expansion
in the presence of viscous forces and estimate the survival probability
in a longitudinally expanding QGP in Sec.III.B. 
Results and discussion will be presented in Sec.IV and finally, 
we conclude in Sec.V.

\section{Equation of state for Strongly-interacting QGP }
In this section, first we briefly discuss the medium modified effective 
potential in Sec.II.A. In Sec.II.B, we revisit the EoS of strongly coupled
QGP developed by Bannur~\cite{banscqgp} and then obtain our EoS
as a function of plasma parameter obtained from the medium-modified potential, discussed in Sec.II.A.

\subsection{Medium modified effective potential}
In thermodynamical studies of QCD plasma phase, deviations from 
perturbative calculations and ideal gas
behavior at temperatures much larger than the deconfinement temperature 
calls for quantitative non-perturbative calculations~\cite{kaczmarek}.
In light of this finding, one can not simply 
ignore the effects of string tension between the quark-antiquark pairs
beyond $T_c$. Recently, this issue had successfully been addressed in 
the context of the dissociation of quarkonium in 
QGP~\cite{prc-vineet,chic-vineet} where we assumed the potential between
a heavy quark-antiquark at $T=0$ as the Cornell potential
: $V(r)=-\alpha/r +\sigma r$ and then corrected its
Fourier transform (FT) $\tilde{V} (k)$, to incorporate the medium 
modifications, as 
\begin{equation}
\label{eq3}
\tilde{V}(k)=\frac{V(k)}{\epsilon(k)} \quad ,
\end{equation}
where $\epsilon(k)$ is the dielectric permittivity~\cite{schneider}, given
by
\begin{eqnarray}
\label{eqn4}
\epsilon(k)=\left(1+\frac{ \Pi_L (0,k,T)}{k^2}\right)\equiv
\left( 1+ \frac{m_D^2}{k^2} \right)~,
\end{eqnarray}
and $V(k)$ is the Fourier transform (FT) of
the Cornell potential:
\begin{equation}
\label{eqn5}
{V}(k)=-\sqrt{(2/\pi)} \frac{\alpha}{k^2}-\frac{4\sigma}{\sqrt{2\pi} k^4}.
\end{equation}
Substituting Eqs.(\ref{eqn4}) and (\ref{eqn5}) into (\ref{eq3})
and then evaluating its inverse FT
one obtains the r-dependence of the medium modified
potential~\cite{prc-vineet,chic-vineet},
\begin{eqnarray}
\label{full}
{V}(r)&=& \left( \frac{2\sigma}{m^2_D}-\alpha \right)\frac{e^{-m_D~r}}{r}\nonumber\\
&-&\frac{2\sigma}{m^2_Dr}+\frac{2\sigma}{m_D}-\alpha~m_D~,
\end{eqnarray}
where constant terms arise from the basic computations of real-time 
static potential~\cite{const1} or from real- and imaginary-time
correlators in a thermal QCD medium\cite{const2} and 
are introduced to yield the correct limit of $V(r,T)$ as $T\rightarrow 0$.
The potential (\ref{full}) thus obtained has an additional 
long range Coulomb term with an (reduced) effective charge in addition
to the conventional Yukawa term. 
It is worth to mention~\cite{dixit} that one-dimensional Fourier
transform of the
Cornell potential in the medium yields the screened form as
used in the lattice QCD to study the quarkonium properties
which assumes the one-dimensional color flux tube structure.
However, at finite temperature that may not be the case since
the flux tube structure may expand in more dimensions \cite{satz}.
Therefore, it is better to consider the three-dimensional form
of the medium modified Cornell potential.

Recently we had employed the above medium-modified potential to estimate 
the dissociation pattern of
the charmonium and bottomonium states and also explore how the
pattern changes as we go from perturbative to nonperturbative 
regime~\cite{prc-vineet}. The results obtained in 
perturbative to nonperturbative domain were 
closer to the results obtained from the study of spectral function constructed in lattice QCD and potential model studies, respectively.
This medium modified potential will be employed 
to develop the equation of state in next section.

\subsection{Equation of state}
The equation of state for the quark matter produced in 
relativistic nucleus-nucleus collisions is an important observable and
the properties of the matter are sensitive to it. The expansion of QGP is
quite sensitive to EoS through the speed of sound 
which, in turn, explores the sensitivity of the quarkonium suppression to the equation of state~\cite{dpal,compare}. 

Lattice results~\cite{leos} show that bulk thermodynamic observables
such as pressure and energy density deviate from ideal
gas behavior even at $\sim$ 4 $T_c$ and approach the ideal limit
very slowly. This suggest that there are still (nonperturbative) 
interactions present in
the deconfined medium. There have been many attempts to explain this strongly
interacting matter formed during ultrarelativistic collisions.
Many models, such as bag model\cite{rh.1}, quasi-particle model
\cite{ban2,ban3}, strongly interacting quark-gluon plasma (QGP) model
\cite{sqgp} were used to explain the
complicated matter formed. None of the models were fully
accepted by the physics community even though some models
were successful to a great extent. 
An equation of state for such strongly interacting matter 
was developed by treating QGP near and above $T_c$ in Cornell potential
as Coulomb plus linear confinement term 
and compared with the lattice results for pure gauge, two-flavor,
and three-flavor QGP \cite{ka.2}.
Arnold and Zhai \cite{arnold} and Zhai and Kastening \cite{zhai} 
have developed a perturbative EoS of interacting quarks
and gluons upto O($g^5$) which has further been extended 
by Kajantie et al.~\cite{klaine,klaine1} by
including the nonperturbative contributions {\em viz.}
$O(gT)$ and $O(g^2 T)$ up to O($g^6\ln(1/g)$)\cite{krebh}. 
      On the other hand, a semiclassical approach is devised
to study the bulk properties of QGP 
\cite{nayak,bhalo,jain,pfk,pfk1} where 
nonperturbative effects 
manifested as effective mean color fields. These color fields have 
the dual role of producing
the soft and semisoft partons, apart from modulating their 
interactions. The emergence
of such effective field degrees of freedom, together with a 
classical transport has been indicated earlier~\cite{nayak,bhalo}.

Recently Bannur~\cite{banscqgp} developed an equation of state 
for a strongly-coupled QGP by appropriate modifications 
of strongly-coupled plasma in QED to take account color and 
flavor degrees of freedom with the running coupling constant 
and a reasonably good fit to the lattice results 
was obtained. Let us briefly discuss the strongly coupled 
plasma in QED where the equation
of state is expressed as a function of plasma parameter
$\Gamma$~\cite{ic.1}:
\begin{equation} 
\epsilon_{{}_{\rm{QED}}} = \left( \frac{3}{2} + u_{ex} (\Gamma) 
\right) \, n \, T \; , \label{eq:scp} 
\end{equation}
where the first term represents the ideal contribution and the deviations
from ideal EoS is given by, 
\begin{equation} 
u_{ex} (\Gamma) = \frac{u_{ex}^{Abe} (\Gamma) + 3 \times 10^3 \, \Gamma^{5.7}  
  u_{ex}^{OCP} (\Gamma) }{1 + 3 \times 10^3 \, \Gamma^{5.7} } \; , 
  \label{eq:uex} 
\end{equation}
where $u_{ex}^{Abe}$ is given by   
\begin{eqnarray} 
u_{ex}^{Abe} (\Gamma) &=& - \frac{\sqrt{3}}{2} \, \Gamma^{3/2} - 3 \, \Gamma^3 
 \left[ \frac{3}{8} \, \ln (3 \Gamma) + \right. \nonumber\\
&&\left. \frac{\gamma}{2} - \frac{1}{3} \right] ~, 
\end{eqnarray}
which was derived by Abe \cite{ab.1} in the formalism of
giant cluster expansion with the Euler constant $\gamma$ 
and is valid for $\Gamma < .1$. The term $u_{ex}^{OCP}$ is
given by
\begin{eqnarray} 
u_{ex}^{OCP} &=& - 0.898004 \Gamma + 0.96786  \Gamma^{1/4} 
      \nonumber\\
&+& 0.220703  \Gamma^{- 1/4} - 0.86097 \quad ,  \label{eq:uo} 
\end{eqnarray}
which was numerically obtained for one 
component plasma and is valid all $\Gamma < 180$~\cite{ic.2}.
In the above equations, the plasma parameter $\Gamma$ is the ratio 
of average potential energy to the average kinetic energy.   

Let us now consider strongly-coupled plasma in QCD
where it was assumed that hadron exists for $T<T_c$ and goes to QGP for 
$T>T_c$. That is, for $T>T_c$, it is the strongly interacting 
plasma of quarks and gluons and no hadrons or glue balls because 
it was assumed that confinement interactions due to QCD vacuum 
has been melted~\cite{banscqgp} at $T=T_c$.
Hence the only interaction present in the deconfined plasma phase 
is the Coulomb interaction and so the plasma
parameter ($\Gamma$) was evaluated as the ratio of average Coulomb potential
energy to average kinetic energy.
As discussed earlier in Sec.II.A, we will retain confinement 
interactions, in addition to the Coulomb interaction, through the
linear term in the potential which manifests in the nonzero 
values of string tension even at $T \ge T_c$ in the potential (\ref{full}). 
Finally, the equation of state has been obtained by using the
potential (4) in the plasma parameter after inclusion of 
relativistic and quantum effects as: 
\begin{equation} 
\varepsilon = \left( \frac{}{} 3 + u_{ex} (\Gamma) \right) 
\, n \, T \;, \end{equation}
where the form of $u_{ex} (\Gamma)$ remains the same as in (6). In terms 
of ideal contribution, the scaled-energy density is written as
\begin{equation} 
e(\Gamma) \equiv \frac{\varepsilon}{\varepsilon_{SB}} 
= 1 + \frac{1}{3} u_{ex} (\Gamma) \quad,
\end{equation}
where $\varepsilon_{{}_{SB}}$ is given by,
\beq
\varepsilon_{{}_{SB}} \equiv 3 a_f T^4, 
\eeq
where $a_f \equiv (16 + 21 \, n_f /2) \pi ^2 /90$ is a constant 
which depends on degrees of freedom of quarks and gluons. 
Here we have employed the QCD running coupling in $\msbar$ scheme~\cite{shro}, 
in compatible with lattice simulation, up to two-loop level:
\bqa
\label{eqg}
{g^2} \approx 2 b_0 \ln \frac{\bar\mu}{\lambdamsbar}
{\left( 1 + \frac{b_1}{2b_0^2} \frac{\ln \left( 2 \ln 
\frac{\bar \mu}{\lambdamsbar}\right)}{\ln \frac{\bar\mu}{\lambdamsbar}}
\right)}^{-1} \quad ,
\eqa
where $b_0= (33-2n_f)/(48 \pi^2)$ and $b_1= (153-19n_f)/(384\pi^4)$,
$\bar \mu$ and $\lambdamsbar$ are the scale parameter and
the renormalization scale in $\msbar$ scheme, respectively. 
For, the EoS to depend on the renormalization scale,
the physical observables should be scale independent. 
We circumvent the problem
by trading off the dependence on $\lambdamsbar$
to a dependence on the critical temperature $T_c$.
\bqa
\bar\mu~\exp(\gamma_E+c)&=&\lambdamsbar (T)\nonumber\\
\lambdamsbar (T) \exp(\gamma_E+c)&=&4 \pi \Lambda_T \quad,
\eqa
where $c$ is a constant depending on colors and flavors:
$c=\left( n_c-4n_f \ln 4 \right)/\left(22n_c-n_f \right)$
and $\gamma_E$=0.5772156. There are several ambiguities, associated 
with the renormalization scale $\lambdamsbar$, the scale parameter 
$\bar \mu$ which occurs in the
expression for the running coupling constant $\alpha_s$. This issue
has been discussed well in literature
and a popular way out is the BLM criterion due to Brodsky, Lepage 
and Mackenzie~\cite{shaung}. In this criterion, 
$\lambdamsbar$ is allowed to vary between $\pi T$ and $4 \pi T$~\cite{braten}.
For our purposes, we choose the renormalization scale $\lambdamsbar$
close to the central value $2 \pi T_c$~\cite{vuorinen} for $n_f$=0 and 
$\pi T_c$ for both $n_f$=2 and $n_f$=3 flavors.

It is worth to mention here that if the factor
$\frac{b_1}{2b_0^2} \frac{\ln \left( 2 \ln 
\frac{\bar \mu}{\lambdamsbar}\right)}{\ln \frac{\bar\mu}{\lambdamsbar}}$
is much smaller than 1 ($ \ll 1$) then the
above expression reduces to the expression used in~\cite[Eq.(10)]{banscqgp}, after neglecting the higher order
terms of the above factor.
However, this possibility does not hold good
for the temperature ranges used in the calculation and 
cause an error in coupling which finally makes the difference 
in the results between our model and Bannur model~\cite{banscqgp}.
\begin{figure*}
\vspace{-2mm}
\includegraphics[scale=.35]{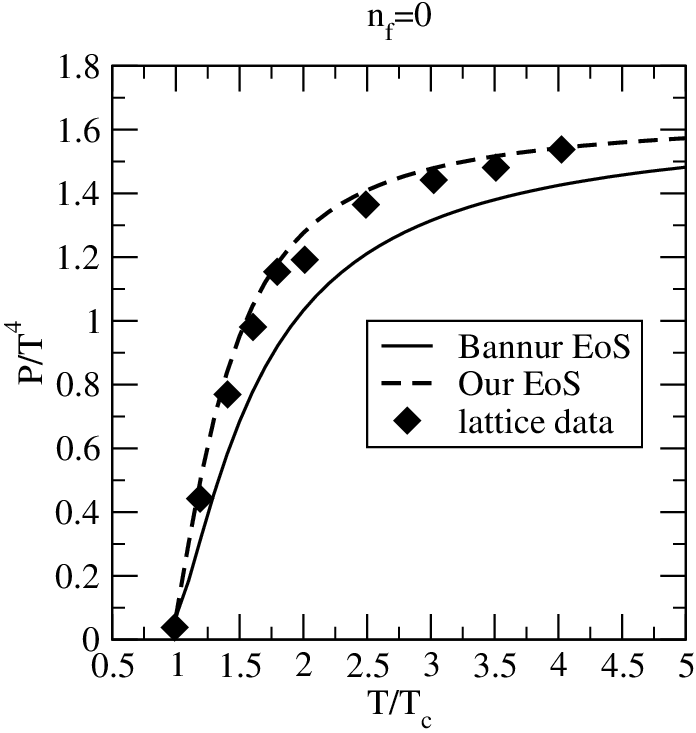}
\hspace{4mm}
\includegraphics[scale=.35]{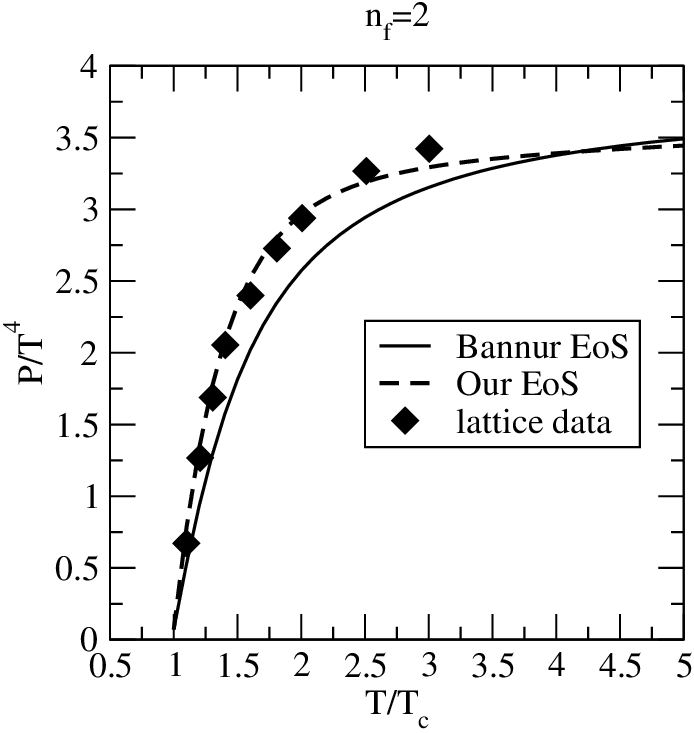}
\hspace{4mm}
\includegraphics[scale=.35]{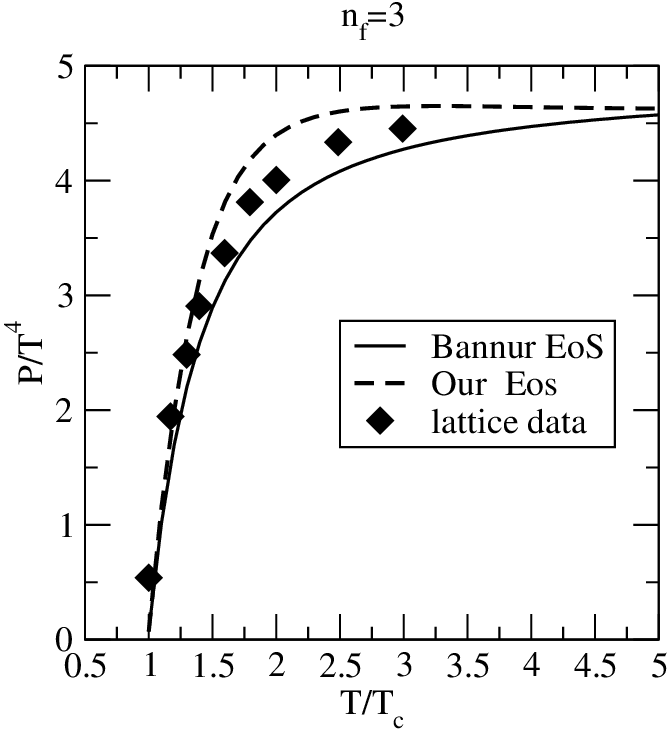}
\vspace{-1mm}
\caption{Plots of $ P/T^4 $ as a function of $T/T_c$ from Bannur EoS, 
our EoS and lattice results for (a) pure gauge 
(extreme left figure), 2-flavor QGP (middle figure) and 3-flavor QGP (extreme 
right figure). In each figure, solid line represents the results obtained 
from Bannur EoS, dashed line represents the results from our EoS and 
diamond symbols represent lattice results.}
\vspace{46mm}
\end{figure*}
Finally, we get the energy density $\varepsilon (T)$ from Eq.(10)and using the 
thermodynamic relation, 
\begin{equation} 
\varepsilon = T \frac{dp}{dT} - P \quad, 
\end{equation}
we get the pressure as 
\begin{equation} \frac{P}{T^4} = \left( \frac{P_0}{T_0} + 3 a_f \int_{T_0}^T \, 
d\tau \tau^2 e(\Gamma(\tau)) \right) / T^3 \; , \label{eq:p} \end{equation} 
where $P_0$ is the pressure at some reference temperature $T_0$
and has been fixed with the values of pressure at critical 
temperature for a particular system - gluon plasma,
2-flavor plasma etc. Once we know the pressure $P$ and energy density 
$\varepsilon$, the speed of sound $c_s^2 (= 
\frac{dP}{d\varepsilon})$ can be evaluated. 

In Fig. 1, we have plotted the variation of pressure ($P/T^4$) with
temperature ($T/T_c$) for
pure gauge, 2-flavor and 3-flavor QGP along with lattice results.
In Bannur model~\cite{banscqgp}, 
for each system, $g_c$ and $\Lambda_T$ are adjusted to get a good fit 
to lattice results. However, in our calculation, there is no quantity 
to be fitted for predicting lattice results.
We have fixed $P_0$ from the lattice data at the critical temperature $T_c$
for each system, separately. 

\begin{figure*}
\vspace{-1mm}
\includegraphics[scale=.40]{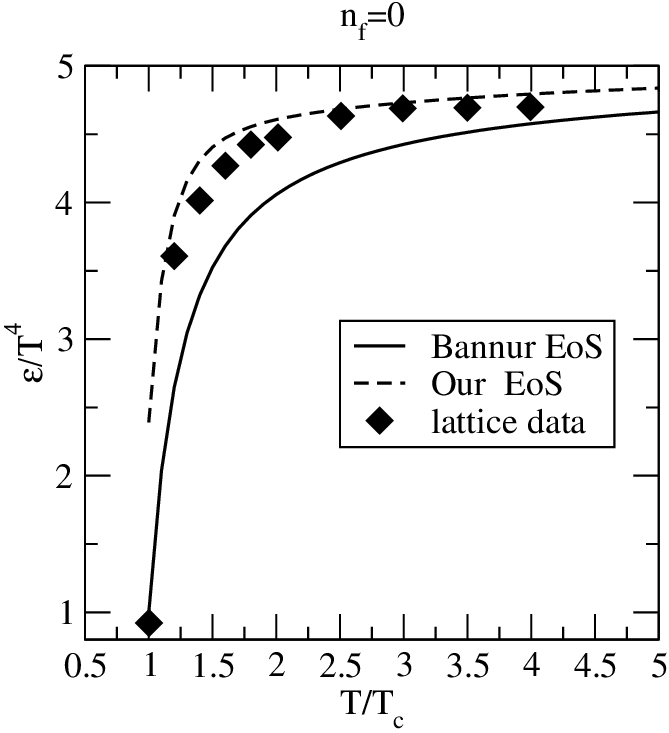}
\hspace{4mm}
\includegraphics[scale=.40]{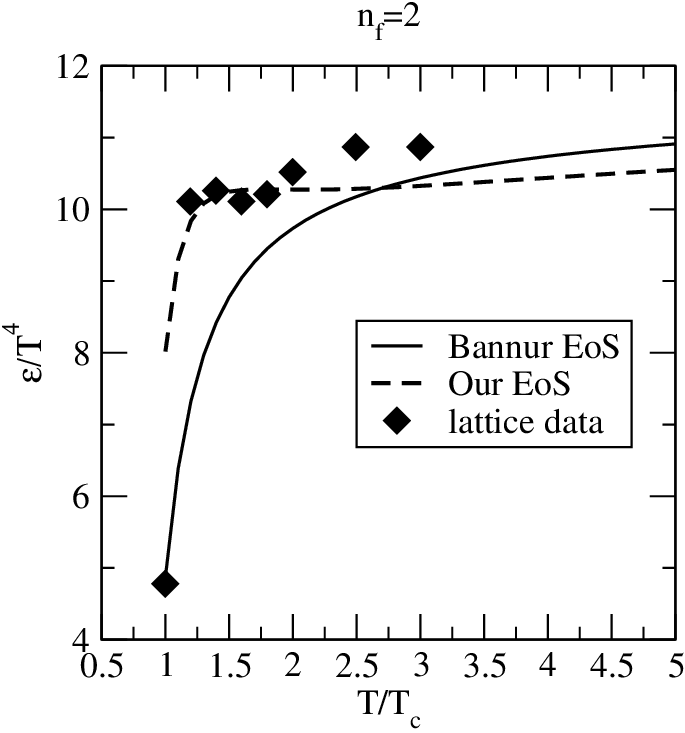}
\hspace{4mm}
\includegraphics[scale=.40]{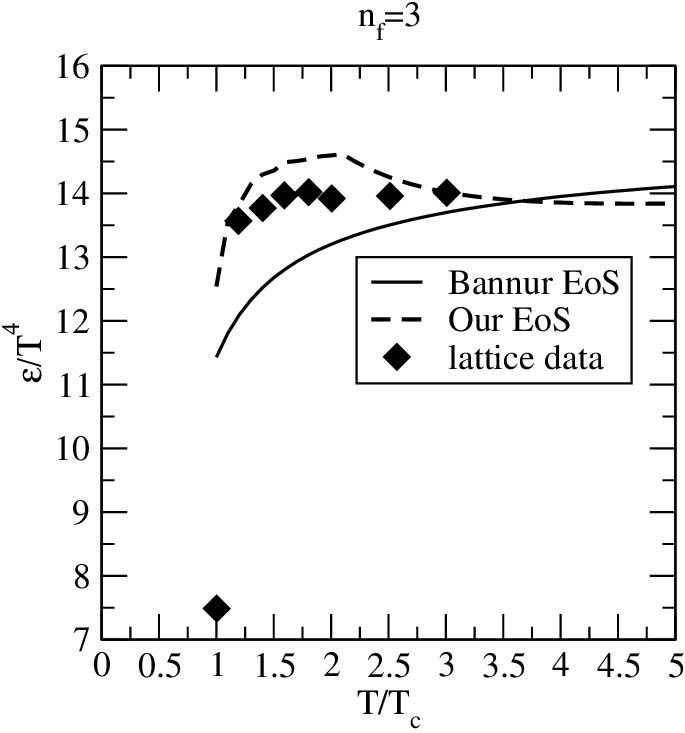}
\vspace{-1mm}
\caption{Plots of $\varepsilon/ T^4 $ as a function of $T/T_c$. The notations
are the same as in Fig.1.}
\vspace{45mm}
\end{figure*}

Once pressure, $P(T)$ is obtained, then other macroscopic quantities such as
energy density $\varepsilon$, speed of sound $c_s^2$ etc. can be derived
from $P(T)$ and no other
parameters are needed. In Fig. 2, we plotted the energy density 
($\varepsilon / T^4$) with temperature ($T/ T_c$) for all three systems 
along with lattice
results and a reasonably good fit is obtained without any extra parameters.
All three curves looks similar, but shifts to left
as flavor content increases. For the sake of comparison with the
results of Bannur EoS, we have taken the critical temperatures 
$T_c$ equal to $275$, $175$ and $155$ MeV for gluon plasma, 2-flavor 
and 3-flavor QGP, respectively.

\begin{figure*}
\vspace{-1mm}
\includegraphics[scale=.40]{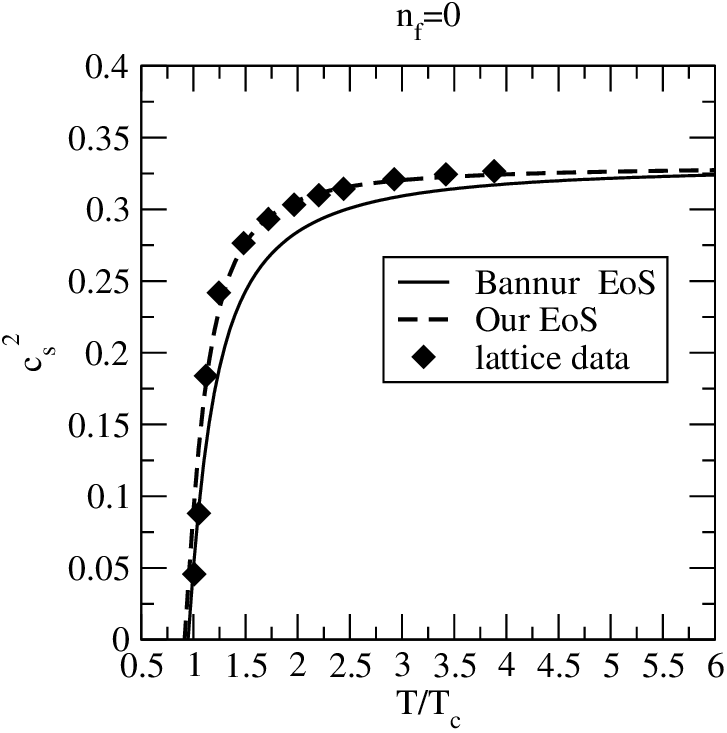}
\hspace{4mm}
\includegraphics[scale=.40]{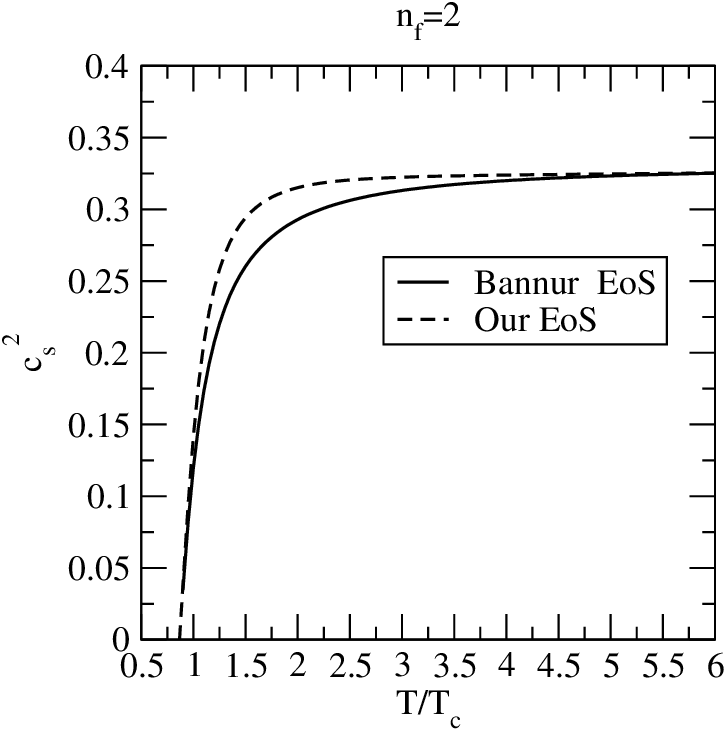}
\hspace{4mm}
\includegraphics[scale=.40]{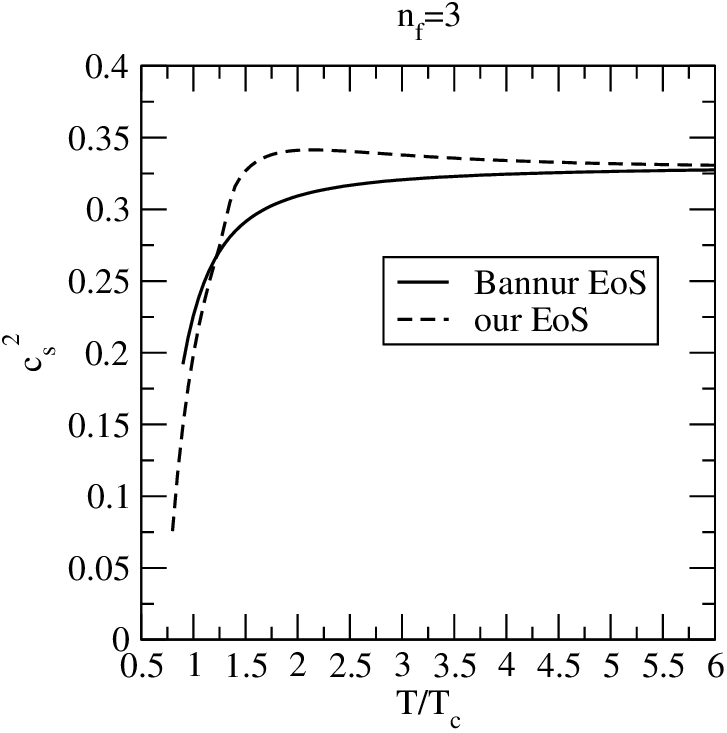}
\vspace{-1mm}
\caption{Plots of $c_s^2$ as a function of $T/T_c$ from 
our EoS and Bannur EoS  for pure gauge,2-flavor QGP and 3-flavor QGP.} 
\vspace{45mm}
\end{figure*}

In Fig. 3, the speed of sound, $c_s^2$ is plotted for all three systems, 
but matching have been checked with the only available lattice results
for gluon plasma. There is an excellent
agreement with the lattice results
and we have predicted the results for the flavored QGP. All three curves have
similar behaviour, i.e, sharp rise near $T_c$ and then flatten to the
ideal value ($1/3$). However, $c_s^2$ is larger 
for larger flavor content of plasma. In the vicinity of critical
temperature, fits or predictions may not be good, especially for
energy density $\varepsilon$ and $c_s^2$ which strongly depends on 
variations of pressure $P$ with respect to temperature $T$. Lattice data 
also has large error bars
very close to $T_c$. However, except for small region at
$T=T_c$, our results are very good for all regions of $T > T_c$.
It is interesting to note that recently
Peshier and Cassing~\cite{pe.3} also obtained similar results on
the dependence of plasma parameter $\Gamma$ in quasi-particle model 
and concluded that QGP behaves like a liquid, not weakly-interacting gas.

Now we are interested to study for the realistic case where u and d quarks
have very small masses (5-10 MeV) and strange quarks are having masses
150-200 MeV and charm quark with mass 1.5 GeV. 
Let $g_f$ counts the effective number of degrees of freedom of a 
massive Fermi gas. For a massless gas we have, of course, $g_f = n_f$.
In general we define
\begin{eqnarray}
n_f &=&\sum_{f=u,d,\cdot \cdot\cdot} g(m_f/T)\\
{\rm{where}},\nonumber\\
g(\frac{m_f}{T} ) &=&\frac{360}{7\pi^4} \int_0^\infty dx~x~
\sqrt{x^2-{ \left(\frac{m_f}{T}\right)}^2}~\times \nonumber\\ 
&&\ln \left( 1 + e^{-x} \right) 
\end{eqnarray}

In Fig.4, we have shown our results on (2+1)-flavors and 4-flavors QGP
along with lattice data~\cite{ka.3,ka.4} and
replotted the variation of $P(T)/T^4$ with temperature $T/T_c$ for all systems.
Similar plots for energy density $\varepsilon (T)/T^4$ with 
temperature $T/T_c$ for all systems is replotted in Fig. 5. 
We have also compared with the resuts from Bannur model.

\begin{figure*}
\vspace{-1mm}
\includegraphics[scale=.35]{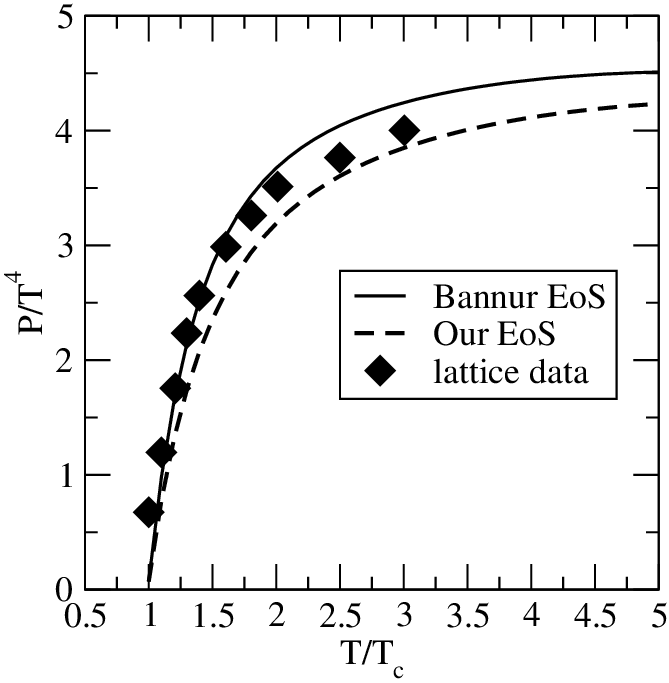}
\hspace{4mm}
\includegraphics[scale=.35]{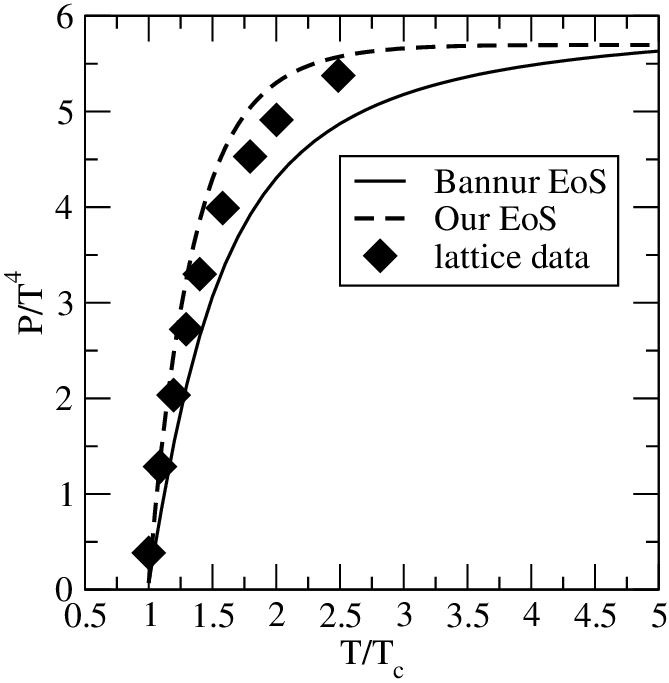}
\hspace{4mm}
\includegraphics[scale=.35]{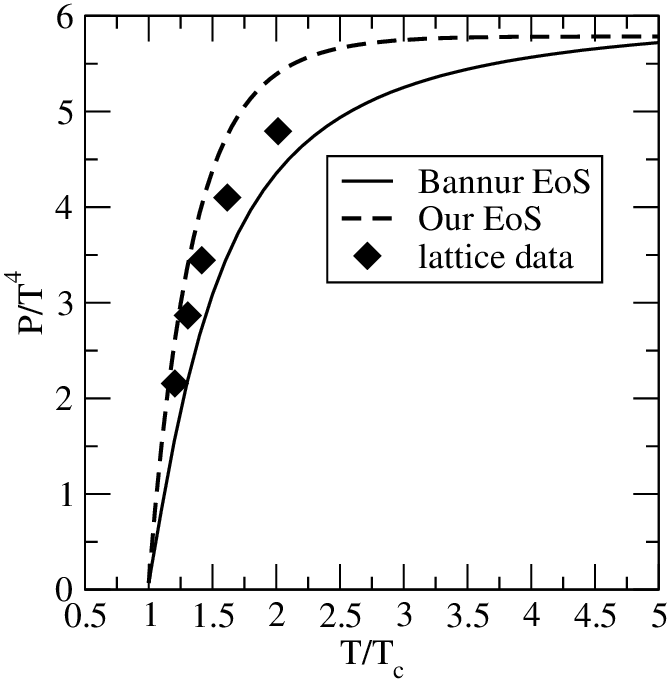}
\vspace{8mm}
\caption{variation of $P/T^4 $ as a function of $T/T_c$ for a) two
massless and one massive (2+1), b) and c) for 4-flavour QGP for 
two different masses, $m/T$=0.4 and 0.2, respectively. The notations 
are the same as in Fig.1.}
\vspace{50mm}
\end{figure*}
\begin{figure*}
\vspace{-2mm}
\includegraphics[scale=.35]{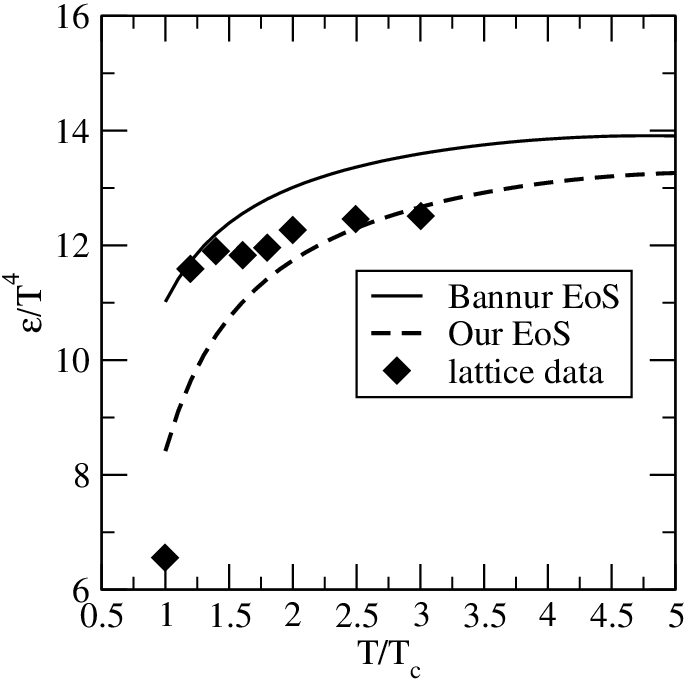}
\hspace{4mm}
\includegraphics[scale=.35]{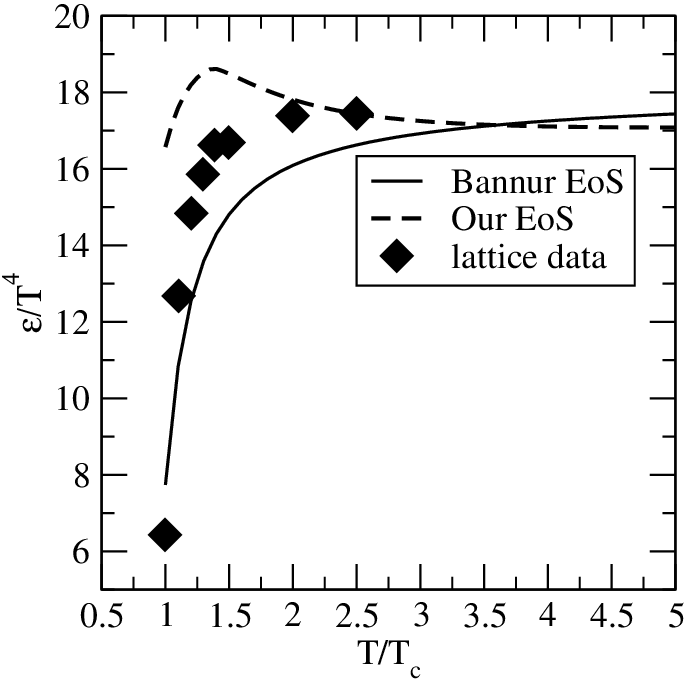}
\hspace{4mm}
\includegraphics[scale=.35]{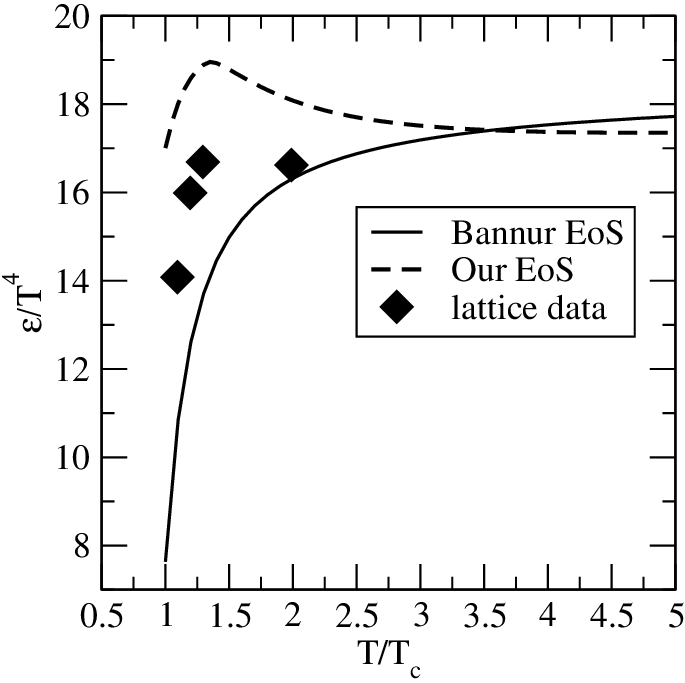}
\vspace{-1mm}
\caption{Variation of $\varepsilon/ T^4 $ as a function of $T/T_c$ 
where the notations are the same as in Fig.4.}
\vspace{25mm}
\end{figure*}

This indicates that in the presence of a heavier quark the 
deviations of the pressure from the ideal gas value is larger 
than in the massless limit. This is in qualitative
agreement with the observations. In view of the agreement with 
the results of lattice equation of state,
our EoS is a right choice for the 
strongly-interacting matter possibly formed at RHIC to calculate the
thermodynamical quantities {\em viz.} screening energy density ($\epsilon_s$),
the speed of sound etc. to study the hydrodynamical expansion of plasma
and finally, to estimate the suppression of $J/\psi$ in nuclear collisions.
\section{Suppression of $J/\psi$ in a longitudinally expanding plasma}
In Sec.III.A, we study hydrodynamic boost-invariant Bjorken expansion in $(1+1)$ dimension with the EoS discussed 
in Sec.IIB as an input. In addition, we explore the 
effects of dissipative terms up to first-order in the stress-tensor. 
Then we turn our attention to derive 
the $J/\psi$ survival probability for an expanding QGP, in Sec.III.B.

\subsection{Longitudinal expansion in the presence of dissipative forces}

In the presence of viscous forces, the energy-momentum tensor is written as,
\begin{equation}
T^{\mu\nu}= (\epsilon+p)u^\mu u^\nu + g^{\mu \nu} p +\pi^{\mu\nu},
\label{tmun}
\end{equation}
where the stress-energy tensor, $\pi^{\mu\nu}$ up to first-order 
is given by
\beq
\pi^{\mu\nu}=\eta \langle \nabla^\mu u^\nu \rangle ~, 
\eeq
where $\eta$ is the co-efficient of the shear viscosity and 
$ \langle \nabla^\mu u^\nu \rangle$ is the symmetrized velocity gradient. 

In (1+1) dimensional Bjorken expansion in the first-order
dissipative hydrodynamics, only one component $\pi^{\eta \eta}$ of the 
viscous stress tensor is non-zero, hence the equation of motion reads,
\beq
\label{eqm}
\partial_\tau \epsilon =-\frac{\epsilon+p}{\tau}+\frac{4 \eta}
{3 \tau^2}\, .
\label{3.15}
\eeq
The first term in the RHS is the same as in the case of
zeroth-order (non-viscous) hydrodynamics and the second term is the
correction arising from constant $\eta/s$ which causes the system
to expand slowly compared to the perfect fluid, $\eta=0$.

The solution of equation of motion (19) is obtained as,
\begin{eqnarray}
\label{eqs1}
\epsilon(\tau) \tau^{(1+c_s^2)}+ \frac{4a}{3{\tilde{\tau}}^2}
\tau^{(1+c_s^2)}
&=&\epsilon(\tau_i)\tau_i^{(1+c_s^2)}
+\frac{4a}{3 {\tilde{\tau_i}}^2} \nonumber\\
&=&{{\mbox{const}}}~,
\end{eqnarray}
where the constant $a$ is $\left(\frac{\eta}{s}\right)T^3_i \tau_i$ and 
the symbols, ${\tilde{\tau}}^2$ and ${\tilde{\tau}}_i^2$ are given by
$(1-c_s^2)\tau^2$ and $(1-c_s^2)\tau_i^2$, respectively.
The first term accounts for the contributions coming from the 
zeroth-order expansion (ideal fluid) and the second term is the first-order 
viscous corrections. The above equation (21) will finally be coupled with
the dissociation of a quarkonium in a static thermal medium to estimate
the quarkonium suppression in relativistic nucleus-nucleus collisions.

\subsection{Survival probability}
We now have all the ingredients to write down the survival probability.
Chu and Matsui~\cite{chu} studied 
the transverse momentum dependence ($p_T$) of the survival probability
by choosing the speed of sound $c_s^2=1/3$ (ideal EoS)
and the extreme value $c_s^2=0$.
This work was further generalized by invoking the
various parameters for Au-Au collisions at RHIC in~\cite{dpal}
to include the effects of realistic EoS in
an adhoc manner by simply choosing a lower value of $c_s^2=1/5$.
Instead of taking arbitrary values of $c_s^2$ we tabulated
the values of $c_s^2$ in Tables II and IV-VI corresponding to the
dissociation temperatures calculated from our EoS~\cite{eurjc-vineet}
for charmonium and bottomonium states, respectively.
Moreover, in the light of recent experimental finding from RHIC,
one cannot ignore the viscous effects while studying
charmonium suppression. Here, we address these issues.

Let us take an initial energy-density profile on a transverse plane :
\begin{equation}
\epsilon(r;\tau_i)=\epsilon_i \left(1-
\frac{r^2}{R_T^2}\right)^{\beta} \Theta(R_T-r)
\end{equation}
where $r$ is the transverse co-ordinate and $R_T$ is the transverse 
radius of the nucleus.
One can define an average energy density $\langle \epsilon_i \rangle $ as
\begin{equation}
\pi R_T^2 \langle \epsilon_i \rangle =\int 2\pi\, r \, dr \epsilon(r;\tau_i)
\end{equation}
so that 
\begin{equation}
\epsilon_i=(1+\beta)  \langle \epsilon_i \rangle ~~~;\beta=1.
\end{equation} 
The average initial energy density ${\langle \epsilon_i \rangle}$~\cite{khar}
will be given by the modified Bjorken formula:
\begin{equation}
{\langle \epsilon_i \rangle}=\frac{\xi}{A_T\,\tau_i}
\left(\frac{dE_T}{dy_h}\right)_{y_h=0} ,
\end{equation}
where $A_T$ is the transverse overlap area of the colliding nuclei and 
$(dE_T/dy_h)_{y_h=0}$ is the transverse energy deposited per unit 
rapidity of outgoing hadrons. Both depend on the number of participants 
$N_{part}$~\cite{sscd} and thus provide centrality dependent initial 
average energy density ${\langle \epsilon_i \rangle}$ in the transverse 
plane (Table I). For this purpose, we have extracted the transverse overlap area
$A_T$ and the pseudo-rapidity distribution ${dE_T/d\eta_h \mid}_{\eta_h=0}$
~\cite{sscd} at various values of  number of participants
$N_{part}$. These ${dE_T/d\eta_h \mid}_{\eta_h=0}$ numbers are then multiplied
by a Jacobian 1.25 to yield the rapidity distribution
${dE_T/dy_h \mid}_{y_h=0}$ which will be further used to calculate
the average initial energy density from Bjorken formula (25).
The scaling factor $\xi=5$ has been introduced in order to obtain the 
desired values of initial energy densities  
~\cite{hiran} for most central collision
which are consistent with the predictions of the self-screened parton 
cascade model~\cite{eskola} and also with the 
requirements of hydrodynamic simulation~\cite{hiran} to fit
the pseudo-rapidity distribution of charged particle 
multiplicity $dN_{ch}/d\eta$ for various centralities 
observed in PHENIX experiments at RHIC energy. Later we will discuss
the centrality dependence of initial energy densities at LHC energy
in Sec.IV.
\begin{table}[tbp]
\caption{Kinematic characterization of Au$+$Au collisions at 
RHIC~\cite{expt}}
\begin{center}
\begin{tabular}{lll|lll}
\hline Nuclei & $\sqrt{s_{NN}}$ & $N_{\mbox{part}}$~ &~ ${\langle \epsilon_i 
\rangle}$~
~ &~ $R_T$ \\ 
& (GeV) & ~ &~ (GeV/fm$^3$)~ &~ (fm)\\ \hline
         &          & 22.0 & 5.86 & 3.45 \\
         &          & 30.2 & 7.92 & 3.61 \\
         &          & 40.2 & 10.14& 3.79 \\
         &          & 52.5 & 12.76& 3.96 \\
         &          & 66.7 & 15.69& 4.16 \\
         &          & 83.3 & 18.58& 4.37 \\
Au$+$Au  & 200  & 103.0& 21.36& 4.61 \\
         &          & 125.0& 24.38& 4.85 \\
         &          & 151.0& 27.37& 5.12 \\
         &          & 181.0& 30.52& 5.38 \\
         &          & 215.0& 34.17& 5.64 \\
         &          & 254.0& 37.39& 5.97 \\
         &          & 300.0& 41.08& 6.31 \\
         &          & 353.0& 45.09& 6.68 \\\hline
\end{tabular}
\end{center}
\end{table}

The (screening) time, $\tau_s$ when the energy density of the system 
drops to the screening energy density $\epsilon_s$ is estimated 
from Eq.({\ref{eqs1}}) as
\begin{eqnarray}
\label{taus}
\tau_s(r)=\tau_i {\bigg[ \frac{\epsilon_i(r)-
\frac{4a}{3{\tilde{\tau}}_i^2}}{\epsilon_s-\frac{4a}{3{\tilde{\tau}}_s^2}}
\bigg]}^{\frac{1}{1+c_s^2}}
\end{eqnarray}
where $\epsilon_i(r) \equiv \epsilon(\tau_i;r)$ and 
${\tilde{\tau}}_s^2$ is $(1-c_s^2)\tau_s^2$.
The critical radius $r_s$, is seen to mark the boundary of the region 
where the quarkonium formation is suppressed, can be obtained by
equating the duration of screening
$\tau_s(r)$ to the formation time $t_F=\gamma \tau_F$ for the quarkonium
in the plasma frame and is given by:
\begin{eqnarray}
\label{rs}
r_s= R_T { \left( 1- A \right)}^{1/2}~\Theta \left( 1-A \right)~,
\end{eqnarray}
where $A$ is given by
\begin{eqnarray}
A &=& \bigg[  \bigg( \frac{\epsilon_s}{\epsilon_i} \bigg)
\bigg(\frac{t_F}{\tau_i} \bigg)^{1+c_s^2} 
+ \frac{1}{\epsilon_i} {\bigg( \frac{t_F}{\tau_i} \bigg)}^{(1+c_s^2)}
\frac{4a}{3{\tilde{t}}_F^2} \bigg. \nonumber\\
&&+ \bigg. \frac{1}{\epsilon_i} 
\frac{4a}{3{\tilde{\tau}}_i^2} \bigg]^{1/\beta}
\end{eqnarray}
with ${\tilde{t}}_F^2= (1-c_s^2)t_F^2$.
The quark-pair will escape the screening region and form quarkonium 
if its position vector $\mathbf{r}$ and 
transverse momentum $\mathbf{p}_T$ are such that
\begin{equation}
\left| \mathbf{r}+\tau_F \mathbf{p}_T/M\right| \geq r_s.
\end{equation}
Thus, if $\phi$ is the angle between the vectors $\mathbf{r}$ and 
$\mathbf{p}_T$, then the above condition reduces to
\begin{equation}
\cos \phi\,\geq\,\left[(r_s^2-r^2)\,M-\tau_F^2\,p_T^2/M\right]/
\left[2\,r\,\tau_F\,p_T\right],
\label{phi}
\end{equation}
which leads to a range of values of $\phi$ when the quarkonium would
escape. Now we can write for the survival probability of the quarkonium:
\begin{eqnarray}
S(p_T)&=&\left[\int_0^{R_T} \, r \, dr \int_{-\phi_{\mbox{max}}}
^{+\phi_{\mbox{max}}}\,
d\phi\, P(\mathbf{r},\mathbf{p}_T)\right]/\nonumber\\&&
\left[2\pi \int_0^{R_T} \, r\, dr\, P(\mathbf{r},\mathbf{p}_T)\right],
\label{spt}
\end{eqnarray}
where $\phi_{\mbox{max}}$ is the maximum positive angle 
($0\leq \phi \leq \pi$)
allowed by Eq.(\ref{phi}):
\begin{equation}
\phi_{\mbox{max}}=\left\{ \begin{array}{ll}
\pi & \mbox{if $y\leq -1$}\\
\cos^{-1} |y| & \mbox{if $-1 < y < 1$}\\
 0          & \mbox{if $y \geq 1$}
 \end{array}
 \right .,
\end{equation}
where
\begin{equation}
y= \left[(r_s^2-r^2)\,M-\tau_F^2\,p_T^2/M\right]/
\left[2\,r\,\tau_F\,p_T\right],
\end{equation}
and $P$ is the probability for the quark-pair production at
($\mathbf{r}$, $\mathbf{p}_T$), in a hard collision which
may be factored out as 
\begin{equation}
P(\mathbf{r},\mathbf{p}_T)=f(r)g(p_T),
\end{equation}
where we take the profile function f(r) as
\begin{equation}
f(r)\propto \left[ 1-\frac{r^2}{R_T^2}\right]^\alpha \Theta(R_T-r)
\end{equation}
with $\alpha=1/2$. 

Often experimental measurement of survival probability at a given
number of participants ($N_{{}_{\rm part}}$) or rapidity ($y$)
is reported in terms of the $\pt$-integrated yield ratio 
(nuclear modification factor) over the range $\ptmin \le \pt \le \ptmax$
whose theoretical expression would be
\begin{equation}
\sinpt  = 
\frac{\int_{\ptmin}^{\ptmax}  d \pt S(\pt)}
{\int_{\ptmin}^{\ptmax} d \pt}
\end{equation}

In nucleus-nucleus collisions, it is known that only about
60\% of the observed $J/\psi$ originate directly in hard collisions while
30\% of them come from the decay of $\chic$
and 10\% from the decay of $\psi^\prime$. Hence, the $\pt$-integrated inclusive
survival probability of $J/\psi$ in the QGP becomes~\cite{satz,dpal}.
\begin{equation}
\langle S^{{}^{\rm incl}} \rangle = 0.6 {\sdir}_{{}_\psi}
+0.3 {\sdir}_{{}_{\chic}}
+0.1 {\sdir}_{{}_{\psi^\prime}}
\end{equation}

\section{Results and discussions}
Before displaying the results, let us discuss the physical understanding
of quarkonium suppression due to screening in the deconfined medium
produced in relativistic nucleus-nucleus collisions. This involves a competition
of various time-scales involved in an expanding plasma. 
First one is the screening time, $\tau_s$ as the time available for the
hot and dense expanding system during which $\jsi$'s are suppressed.
Second one is the formation time of $\jpsi$ in the plasma frame
($t_F=\gamma \tau_F$) which depends on the transverse momentum 
by which the $c \bar c$ pairs was originally produced.
Third one is the cooling rate which depends on the speed of 
sound, $c_s^2$ through the equation of state.
The screening time not only depends upon the 
screening energy density, $\epsilon_s$ but also depends
on the speed of sound through equation of state. The value of 
$\epsilon_s$ is different for different charmonium 
states and is calculated from the equation of state and hence 
varies from one EoS to other. 
If $\eps \gtrsim \epsilon_i$, initial energy density, then 
there will be no suppression at all i.e., 
survival probability, $\spt$ is equal to 1.

More precisely, the screening time depends upon i) the screening energy
density and the difference between a given initial energy 
density $\epsilon_i$ and screening energy density $\eps$: the more will be the difference the more will be the suppression,
ii) the speed of sound:  for $c_s^2$ less than the ideal limit (1/3), the rate
of cooling will be slower which, in turn, makes the 
screening time larger for a fixed difference in ($\epsilon_i$- $\eps$)
and leads to more suppression, and iii) the $\eta/s$ ratio: 
an additional handle to explore the equation of state by controlling
the expansion of the plasma. If the
ratio is non-zero then the cooling will be slower compared to
$\eta/s$=0, so the system
will take longer time to reach $\eps$ resulting the higher value of 
screening time and hence more suppression
compared to $\eta/s=0$. With this physical understanding we analyze 
our results,$\sinpt$ as a function of the number of participants 
$N_{{Part}}$ in an expanding QGP. 

In our analysis, we have employed the
dissociation temperatures for the charmonium states ($J/\psi$, $\chic$ 
etc.) computed from the lattice QCD correlator
studies~\cite{Dat04} in Table II.
The corresponding values of screening energy densities, $\eps$ and 
the speed of sound $c_s^2$ calculated 
in our EoS are also listed which will be used as
inputs along with the kinematic data in Table I, to calculate $\sinpt$.
\begin{table}
\label{table1}
\centering
\caption{Formation time (fm), dissociation temperature
$T_D$~\cite{satz06},
the speed of sound $c_s^2$ and the screening energy density
$\eps$ ($GeV/fm^3$) for charmonium states, calculated both in 
our and Bannur EoS, respectively.}
\vspace{3mm}
\begin{tabular}{|l|l|l|l|l|l|l|}
\hline
State &$\tau_F$  &$T_D$  &$c_s^2$(our) & $c_s^2$(BAN) & $\epsilon_s$(our) & $\epsilon_s$(BAN) \\
\hline\hline
$J/\psi$ &0.89&  2.1 & 0.308 & 0.275 & 29.33 & 32.05 \\
\hline
$\psi'$ & 1.50& 1.12 & 0.255 & 0.214 & 01.94 & 02.36 \\
\hline
$\chic$ &2.00& 1.16 & 0.261 & 0.220 & 02.28 & 0.220 \\
\hline
\end{tabular}
\end{table}
We have shown the variation of $\pt$-integrated survival probability
(in the range allowed by invariant $\pt$ spectrum of $J/\psi$ in
Phenix experiment~\cite{expt}) with 
$N_{{Part}}$ at mid-rapidity in Fig.6.
The experimental data (the nuclear-modification factor $R_{AA}$)
are shown by the squares with error bars whereas
circles represent sequential suppression. It is found that our calculation
matches completely with the experimental results.
To see the importance of confinement interactions in the deconfined
phase, we have also calculated the $\pt$-integrated survival 
probability (denoted by diamonds)
using the equation of state in Bannur model~\cite{banscqgp} where there
are no confinement interactions in the deconfined phase and have not
found the agreement, as seen in our case. This difference is due to
the fact that the screening energy density calculated in our EoS is 
smaller than the value obtained from Bannur model (as seen in Table II).
The smaller value of screening energy density $\epsilon_s$
causes an increase in the screening time and results in 
more suppression to match with the PHENIX results at RHIC.

\begin{figure*}
\vspace{-1mm}
\hspace{4mm}
\includegraphics[scale=.40]{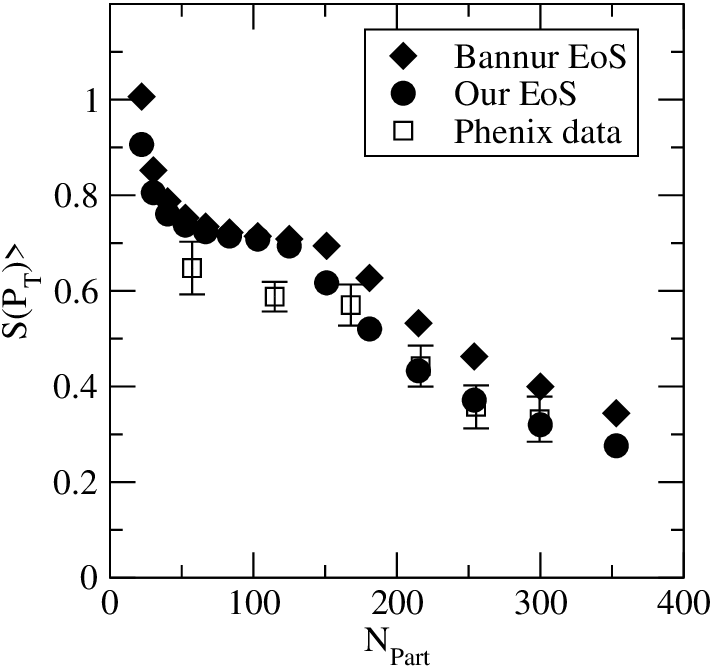}
\hspace{4mm}
\hspace{4mm}
\caption{The variation of $\pt$ integrated survival probability
(in the range allowed by invariant $\pt$ spectrum of $J/\psi$ by the
Phenix experiment~\cite{expt}) versus number of participants at mid-rapidity.
The experimental data (the nuclear-modification factor $R_{AA}$)
are shown by the squares with error bars whereas
circles and diamonds represent sequential melting  using the values of $T_D$'s~
~\cite{prc-vineet} and related parameters from Table II.  }
\vspace{35mm}
\end{figure*}
At RHIC energy, $J/\psi$ yields have been resulted 
from a balance between annihilation of $J/\psi$'s due to hard, thermal 
gluons~\cite{xusat,gluon1} along with colour screening~\cite{mish,chu} 
and enhancement due to coalescence of uncorrelated $c\bar c$ 
pairs~\cite{grand,andro,thews} which are produced thermally at 
deconfined medium, at the phase boundary 
during (statistical) hadronization~\cite{adle,adar}.
However, recent PHENIX data do not show a fully confirmed indication 
of $J/\psi$ enhancement except for the fact that 
$\langle p_T^2\rangle$ of the data and shape of rapidity-dependent 
nuclear modification factor 
$R_{AA}(y)$~\cite{expt} show some characteristics of coalescence production.

On the other hand, at LHC energy, there will be an abundance of thermally 
produced $c \bar c$ pairs so that one cannot make any concrete
conclusion about the possible formation of QGP whereas the number of
$b \bar b$ pairs produced in hot deconfined medium will be meagre 
(because of large bottom quark-antiquark masses) so that the 
competition between the suppression due to screening and dissociation both
and the enhancement due to recombination  may be unlikely.
Therefore, it will be interesting to estimate the $\Upsilon$ suppression
at LHC energy because the initial energy density at ALICE
experiments will be high enough to suppress the 
$\Upsilon$'s  sequentially making the bottomonium suppression
an unambiguous signature.\\

Before finding the centrality (or impact parameter) dependence of
$\Upsilon$ suppression at LHC energy, it is necessary to know the 
centrality dependence of average initial energy density 
${\langle \epsilon_i \rangle}$ in 
terms of the number of participants $N_{part}$ at LHC energy. 
Recently Eskola et al~\cite{eskola} computed the dependence of 
initial energy density, gluon, quark and antiquark numbers produced 
in ultrarelativistic heavy ion collisions with beam energy in 
parton saturation model. This model is 
based on the argument that the effects of all momentum scales can be
estimated by performing the computation at 
the saturation momentum scales. The main
emphasis of the study was at LHC and RHIC energies, although
it gives reasonable good results at SPS too. The dependence
of atomic number and beam energies of initial energy density, number
density, temperature etc. are given by~\cite{eskola}
\bqa
\epsilon_i&=&0.103\,\gev\fm^{-3}A^{0.504}(\roots)^{0.786},\\ 
n_i&=&0.370\,\fm^{-3}A^{0.383}(\roots)^{0.574},\\
T_i&=&0.111\,\gev\, A^{0.126}(\roots)^{0.197}
\label{Tiscaling}
\eqa
Using this model, we have first calculated the 
centrality dependence of the initial energy densities at RHIC energy
and found a good agreement with the known values (Table I)
obtained from the centrality dependence of observed 
particle multiplicities~\cite{sscd}.
Inspired by the success of the parton saturation model at RHIC energy, 
we have provided the centrality dependence of initial conditions at LHC energy
in Table III and predict the suppression of $\Upsilon$ yields at LHC energy,
which is yet to be verified with the results available in ALICE experiments
at LHC energy. 
\begin{table}[tbp]
\caption{Kinematic characterization of Pb$+$Pb collisions at
LHC~\cite{eskola}}
\begin{center}
\begin{tabular}{lll|lll}
\hline Nuclei & $\sqrt{s_{NN}}$ & $N_{\mbox{part}}$~ &~ 
${\langle \epsilon_i \rangle }$~
~ &~ $R_T$ \\
& (GeV) & ~ &~ (GeV/fm$^3$)~ &~ (fm)\\ \hline
         &          & 22.0 & 217.97 & 3.45 \\
         &          & 30.2 & 236.12 & 3.61 \\
         &          & 40.2 & 253.15 & 3.79 \\
         &          & 52.5 & 269.11& 3.96 \\
         &          & 66.7 & 291.45& 4.16 \\
         &          & 83.3 & 315.55& 4.37 \\
Pb$+$Pb  & 5500     & 103.0& 341.05& 4.61 \\
         &          & 125.0& 370.49& 4.85 \\
         &          & 151.0& 400.42& 5.12 \\
         &          & 181.0& 433.18& 5.38 \\
         &          & 215.0& 463.66& 5.64 \\
         &          & 254.0& 507.04& 5.97 \\
         &          & 300.0& 550.88& 6.31 \\

         &          & 353.0& 600.39& 6.68 \\\hline
\end{tabular}
\end{center}
\end{table}
We have taken three sets of dissociation temperatures for the
bottomonium states~\cite{wong76}. These are obtained by equating a) the
lattice free energy, b) by subtracting
entropy term from the lattice free energy and c) linear combination
of both a) and b), with the temperature dependent heavy quark effective
potential.
\begin{table}
\label{table4}\centering
\caption{Formation time (fm), dissociation temperature
$T_D$~\cite{wong76}, the speed of sound $c_s^2$ and the screening 
energy density $\eps$ ($GeV/fm^3$) calculated in our EoS for bottomonium
states, respectively.}
\vspace{3mm}
\begin{tabular}{|l|l|l|l|l|l|l|}
\hline
State &  $\tau_F$  &  $T_D$  &  $c_s^2$(our) & $\epsilon_s$
(our) \\
\hline
$\Upsilon$ &0.76&  4.18 & 0.322 & 496.86 \\
\hline
$\Upsilon'$ & 1.90& 1.47 & 0.287 &  6.31 \\
\hline
$\chi_b$ &2.60& 1.61 & 0.294 & 9.38 \\
\hline
\end{tabular}
\end{table}
\begin{table}
\label{table5}\centering
\caption{Formation time (fm), dissociation temperature
$T_D$~\cite{wong76}, the speed of sound $c_s^2$ and the screening energy density
$\eps$ ($GeV/fm^3$), calculated in our EoS for bottomonium
states,respectively.}
\vspace{3mm}
\begin{tabular}{|l|l|l|l|l|l|l|}
\hline
State &  $\tau_F$  &  $T_D$  &  $c_s^2$(our) & $\epsilon_s$
(our) \\
\hline\hline
$\Upsilon$ &0.76&  3.40 & 0.320 & 213.14 \\
\hline
$\Upsilon'$ & 1.90& 1.18 & 0.263 &  2.44 \\
\hline
$\chi_b$ &2.60& 1.22 & 0.267 & 2.82 \\
\hline
\end{tabular}
\end{table}
\begin{table}
\label{table6}\centering
\caption{Formation time (fm), dissociation temperature
$T_D$~\cite{wong76}, the speed of sound $c_s^2$ and the screening energy density
$\eps$ ($GeV/fm^3$), calculated in SQGP EoS for bottomonium
states,respectively.}
\vspace{3mm}
\begin{tabular}{|l|l|l|l|l|l|l|}
\hline
State &  $\tau_F$  &  $T_D$  &  $c_s^2$(our) & $\epsilon_s$
(our) \\
\hline\hline
$\Upsilon$ &0.76&  2.90 & 0.317 & 111.29 \\
\hline
$\Upsilon'$ & 1.90& 1.06 & 0.244 &  1.47 \\
\hline
$\chi_b$ &2.60& 1.07 & 0.247 & 1.58 \\
\hline
\end{tabular}
\end{table}
In Fig. 7, we have plotted the centrality dependence of the 
$\pt$-integrated survival 
probability for the bottomonium states at LHC energy.
\begin{figure*}
\vspace{-1mm}
\includegraphics[scale=.35]{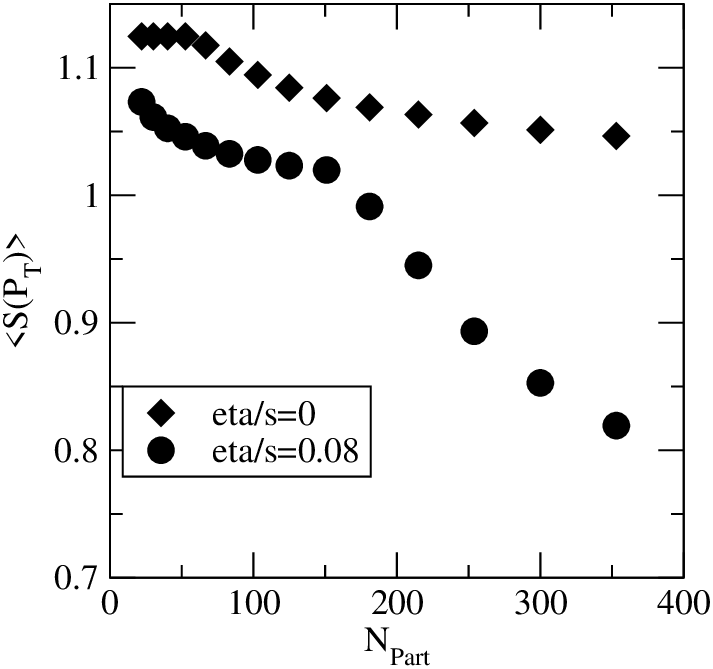}
\hspace{2mm}
\includegraphics[scale=.35]{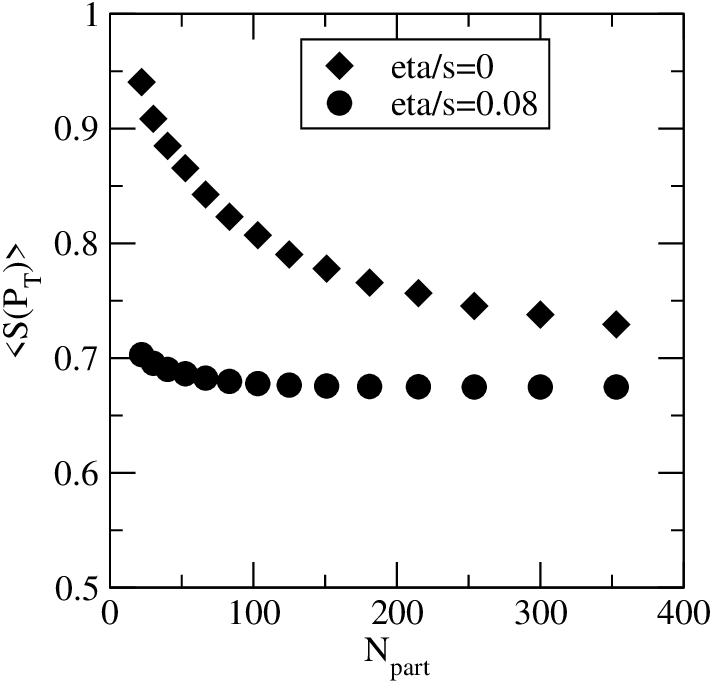}
\hspace{2mm}
\includegraphics[scale=.35]{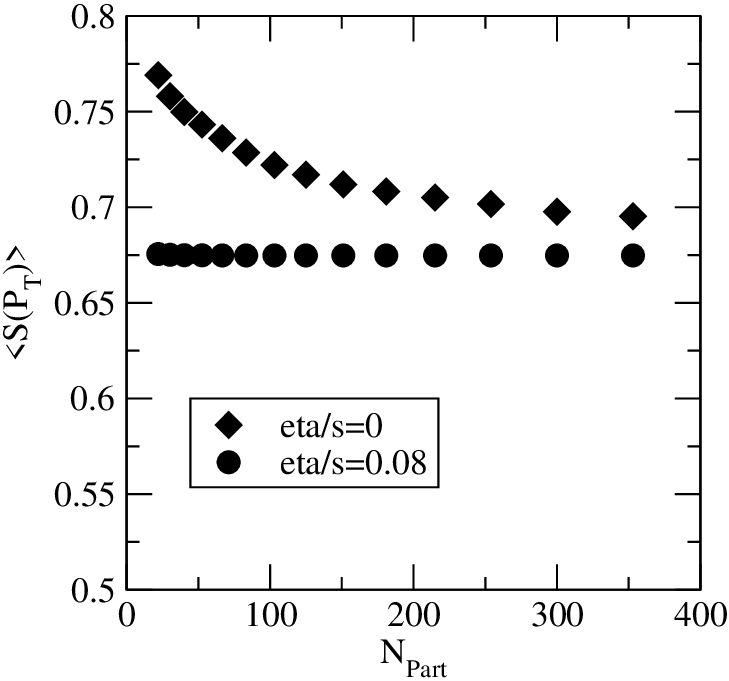}
\vspace{-1mm}
\caption{The variation of $\pt$ integrated survival probability
versus number of participants for $\Upsilon$.The circles 
and diamonds represent sequential melting of $\eta/s=1/4\pi$ 
and $\eta/s=0$, respectively. The parameter for left, middle and right 
figures given in the Table IV,V and VI, respectively.}
\vspace{15mm}
\end{figure*}

\section{Conclusions}
We revisited the equation of state for strongly 
interacting quark-gluon plasma in the framework of strongly coupled plasma 
with appropriate modifications to take account of color and flavor degrees
of freedom and QCD running coupling constant.
In addition, we incorporate the nonperturbative effects in terms 
of nonzero string tension in the 
deconfined phase, unlike the Coulomb interactions alone in the
deconfined phase beyond the critical temperature.
Our results on thermodynamic observables
{\em viz.} pressure, energy density, speed of sound etc. nicely fit the
results of lattice equation of state with gluon, massless and as
well {\em massive} flavored plasma. Motivated by this agreement
we apply our equation of state to estimate the
centrality dependence of $J/\psi$ suppression 
in an expanding dissipative strongly interacting
QGP produced in relativistic heavy-ion collisions.
We have found a complete agreement with the PHENIX experimental results 
on $J/\Psi$ suppression at RHIC energy. Moreover we predicted the same 
for the $\Upsilon$
suppression which is yet to be  verified in ALICE experiments at 
LHC energy.

\end{document}